# A Time Domain Approach to Power Integrity for Printed Circuit Boards

**N. L. Mattey[1]*, G. Edwards[2] and R. J. Hood[2]**

**[1] Electrical & Optical Systems Research Division, Faculty of Engineering, University of Nottingham, University Park, Nottingham, NG7 2RD, UK.**

**[2]Department of Built Environment and Engineering, Deane Building, University of Bolton, Deane Road, BL3 5AB, UK**

*Correspondence to N. L. Mattey, Faculty of Engineering, University of Nottingham, University Park, Nottingham, NG7 2RD, UK. email: N. L. Mattey (nevil.mattey@nottingham.ac.uk)

## Abstract

Power integrity is becoming increasingly relevant due to increases in device functionality and switching speeds along with reduced operating voltage. Large current spikes at the device terminals result in electromagnetic disturbances which can establish resonant patterns affecting the operation of the whole system.

These effects have been examined using a finite difference time domain approach to solve Maxwell's equations for the PCB power and ground plane configuration. The simulation domain is terminated with a uniaxial perfectly matched layer to prevent unwanted reflections. This approach calculates the field values as a function of position and time and allows the evolution of the field to be visualized.

The propagation of a pulse over the ground plane was observed demonstrating the establishment of a complex interference pattern between source and reflected wave fronts and then between multiply reflected wave fronts. This interference which affects the whole ground plane area could adversely affect the operation of any device on the board. These resonant waves persist for a significant time after the initial pulse. Examining the FFT of the ground plane electric field response showed numerous resonant peaks at frequencies consistent with the expected values assuming the PCB can be modelled as a resonant cavity with two electric and four magnetic field boundaries.

## Introduction

Power and signal integrity are of increasing technological relevance due to continuing improvements in speed and functionality of electronic systems and the consequent decreasing operating voltages and increasing magnitude of switching current for typical digital and mixed signal devices. The rapid switching of large numbers of transistors in an integrated circuit causes a large current spike at the power supply terminals resulting in a local ground/power bounce due to parasitic inductances, along with an electromagnetic disturbance to the power planes which may establish a resonant pattern as it is reflected from the edges of the board. This electromagnetic disturbance may compromise the power integrity of other devices or couple to signal traces and affect signal integrity. Here we investigate the effects of board resonance in typical multilayer printed circuit boards (PCBs) by solving Maxwell's equations for the electromagnetic field using the finite difference time domain (FDTD) method. The (FDTD) method gives the electric and magnetic fields as a function of position and time which can then be visualised using animation techniques.

In contrast to frequency domain solutions, a time domain solution captures the response over the bandwidth of the stimulus waveform in a single run and can be more efficient for the large bandwidths resulting from typical clock speeds in digital and mixed signal systems. The evolution of the field distribution can be visualized as the field values are calculated as a function of position and time. Moreover all of the frequency domain information can be extracted via a Fourier transform.

One disadvantage of the FDTD approach is that it is memory intensive as matrices for each dimensional component of the fundamental fields are required for each node of the grid. However continuing advances in computer technologies allow the technique to be applied to realistic models using a standard personal computer (PC).

## Model

Figure 1 shows the modelled configuration where the PCB power and ground planes are separated by a dielectric material and the PCB is surrounded by an air layer.

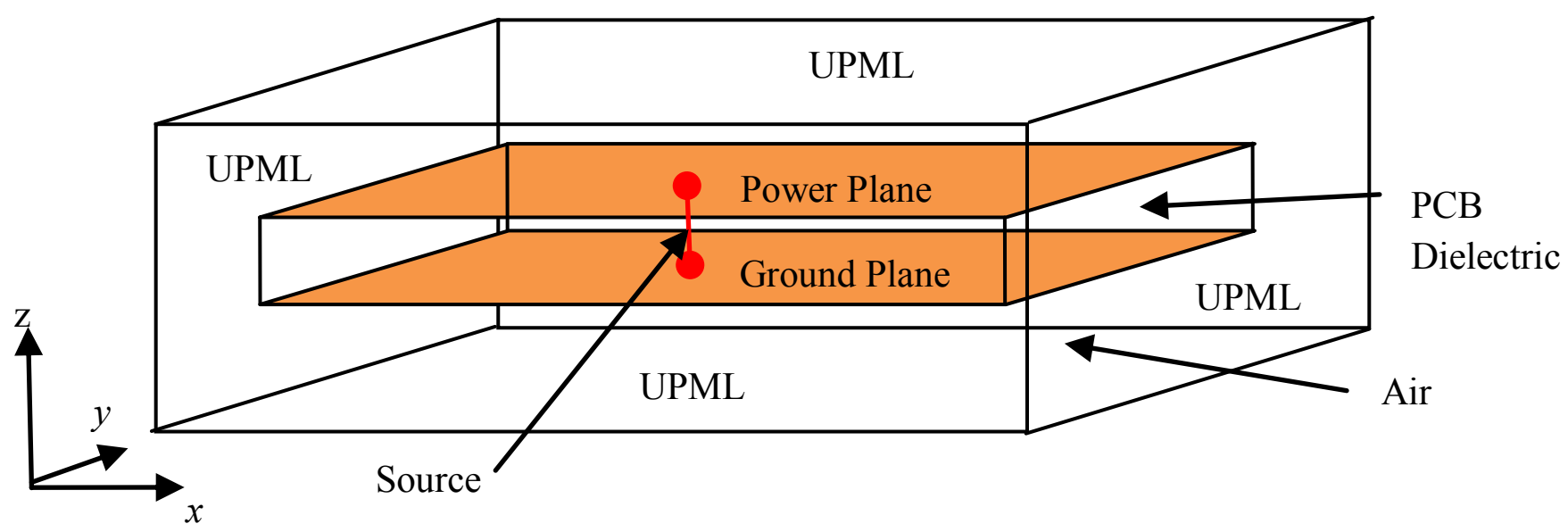

**Figure 1 Modelled PCB configuration surrounded by air gap.**

The model is discretised into a Yee grid as shown in figure 2 where the ***E*** and ***H*** field components are spatially offset so that each ***H*** field component is surrounded by four ***E*** field components and vice versa. In this finite difference time domain (FDTD) scheme [1] the ***E*** and ***H*** field components are also offset by half of a time step. The vector field components at each node of the Yee grid are calculated using explicit or semi-explicit finite centred difference (central-difference) expressions to calculate the space and time derivatives in Maxwell's curl equations based on the previous field vector value, the values of the field vectors at neighbouring nodes (and a known source). The computation begins with all field components set to zero.

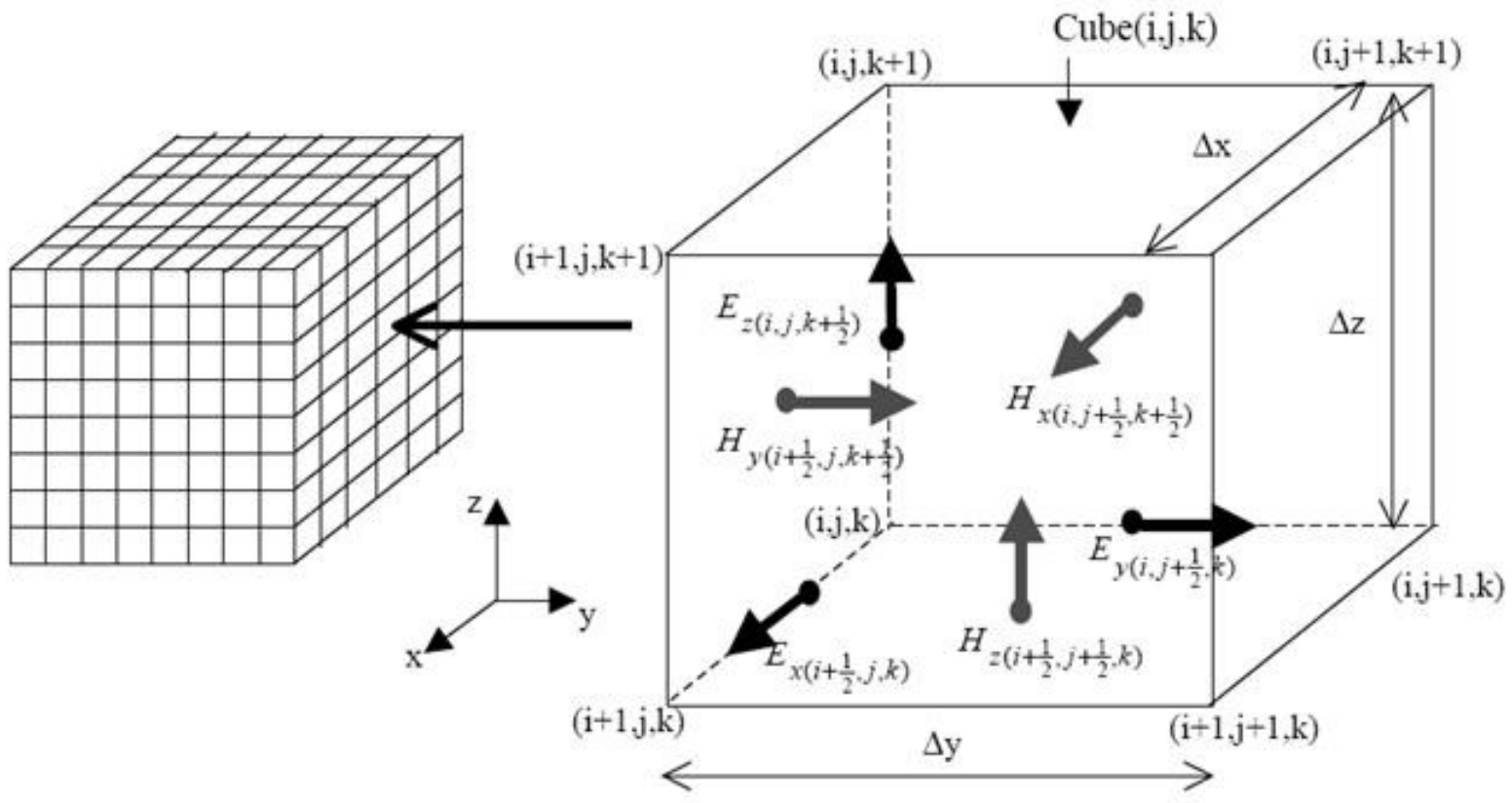

**Figure 2 Yee grid (Taken from [2]).**

The computational domain is terminated by a uniaxial perfectly matched absorbing boundary layer (UPML) [3] whereby plane waves of arbitrary incidence, frequency and polarization are matched at the boundary to a lossy material for inhomogeneous, dispersive, anisotropic or even non-linear domains. Plane waves propagating from the model space decay exponentially within the UPML. This is achieved by defining an artificial anisotropic absorbing material with $\varepsilon$ and $\mu$ tensors as shown in equations 1 and 2 (where the subscripts 1 and 2 represent the model space and the UPML respectively):

$$\bar{\bar{\varepsilon}}_2 = \varepsilon_1 \bar{\bar{s}} \quad \bar{\bar{\mu}}_2 = \mu_1 \bar{\bar{s}} \quad \bar{\bar{s}} = \begin{bmatrix} s_y s_z s_x^{-1} & 0 & 0 \\ 0 & s_x s_z s_y^{-1} & 0 \\ 0 & 0 & s_x s_y s_z^{-1} \end{bmatrix} \tag{1}$$

Where:

$$s_x = \kappa_x + \frac{\sigma_x}{j\omega\varepsilon_0\varepsilon_{rx}} \;,\; s_y = \kappa_y + \frac{\sigma_y}{j\omega\varepsilon_0\varepsilon_{ry}} \;,\; s_z = \kappa_z + \frac{\sigma_z}{j\omega\varepsilon_0\varepsilon_{rz}} \tag{2}$$

This variant of the standard FDTD "leapfrog" algorithm allows a unified treatment of both the model space and the UPML boundaries with $\sigma_w$ and $\kappa_w$ are set to 0 and 1 respectively in the model space. Starting from the time-harmonic form of Maxwell's curl equation for ***H*** i.e

$$\begin{bmatrix} \frac{\partial H_z}{\partial y} & - & \frac{\partial H_y}{\partial z} \\ \frac{\partial H_x}{\partial z} & - & \frac{\partial H_z}{\partial x} \\ \frac{\partial H_y}{\partial x} & - & \frac{\partial H_x}{\partial y} \end{bmatrix} = j\omega \begin{bmatrix} \frac{s_y s_z}{s_x} & 0 & 0 \\ 0 & \frac{s_x s_z}{s_y} & 0 \\ 0 & 0 & \frac{s_x s_y}{s_z} \end{bmatrix} \begin{bmatrix} E_x \\ E_y \\ E_y \end{bmatrix} \tag{3}$$

which can be re-written in terms of ***D*** with $D_x = \varepsilon_0 \varepsilon_r \frac{s_z}{s_x} E_x$, $D_y = \varepsilon_0 \varepsilon_r \frac{s_x}{s_y} E_y$, $D_z = \varepsilon_0 \varepsilon_r \frac{s_y}{s_z} E_z$ .

Substituting for $s$ in terms of $\kappa$ and $\sigma$ and applying the inverse Fourier transform yields a system of time domain differential equations:

$$\begin{bmatrix} \frac{\partial H_z}{\partial y} & - & \frac{\partial H_y}{\partial z} \\ \frac{\partial H_x}{\partial z} & - & \frac{\partial H_z}{\partial x} \\ \frac{\partial H_y}{\partial x} & - & \frac{\partial H_x}{\partial y} \end{bmatrix} = \frac{\partial}{\partial t}\begin{bmatrix} \kappa_y & 0 & 0 \\ 0 & \kappa_z & 0 \\ 0 & 0 & \kappa_x \end{bmatrix}\begin{bmatrix} D_x \\ D_y \\ D_y \end{bmatrix} + \frac{1}{\varepsilon_0}\begin{bmatrix} \sigma_y & 0 & 0 \\ 0 & \sigma_z & 0 \\ 0 & 0 & \sigma_x \end{bmatrix}\begin{bmatrix} D_x \\ D_y \\ D_y \end{bmatrix} \tag{4}$$

Equation 4 can be discretized on the Yee grid giving explicit time-stepping expressions for the components of ***D*** in the UPML from which the ***E*** field components can be calculated in a two step process. A similar two step process can be used for the ***H*** and ***B*** field components. The main disadvantage of this approach compared to the standard method is the need to store the ***D*** and ***H*** vectors in addition to ***E*** and ***H*** which effectively doubles the memory requirements. Although the UPML is theoretically reflectionless, numerical artefacts can occur due to the spatial discretization. To overcome this, $\boldsymbol{\kappa}$ and $\boldsymbol{\sigma}$ are graded over several cells with the outer boundary a perfect electrical conductor. A polynomial grading may be used [3] – for example in the *x* axis:

$$\sigma_x(x) = ({}^{x}/_{d})^m \sigma_{xmax}\,, \quad \kappa_x(x) = 1 + (\kappa_{xmax} - 1)({}^{x}/_{d})^m \tag{5}$$

Thus $\sigma_x$ varies from a value of $0$ at $x = 0$ (the PML boundary) to a value of $\sigma_{xmax}$ at the PEC.

A MATLAB code developed by Willis and Hagness [4] to model a 3D dielectric region with a UPML boundary was modified to model the configuration shown in figure 1. A key modification was to allow an anisotropic grid size to capture the smaller geometry in the *z* axis without the grid becoming excessively large. Note a minimum cell size of one tenth of the shortest wavelength of interest is required to avoid dispersion effects [5] i.e.

$$\Delta x, \Delta y, \Delta z \leq \frac{\lambda_0}{10} \tag{6}$$

In addition the Courant-Freidrichs-Lewy (CFL) stability criterion bounds the time step relative to the grid discretization to ensure numerical stability [5] and is given by:

$$t < \frac{1}{v\sqrt{\frac{1}{\Delta x^2}+\frac{1}{\Delta y^2}+\frac{1}{\Delta z^2}}} \tag{7}$$

where the phase velocity *v* is given by $v = \frac{1}{\sqrt{\mu_0 \mu_r \varepsilon_o \varepsilon_r}}$.

The power and ground planes are modelled as perfect electrical conductors (i.e. the tangential ***E*** field components $E_x$ and $E_y$ are set to zero throughout).

## Results

Figure 3 shows the evolution of the *z* component of the electric field at the ground plane as a function of time where the voltage source is located at the centre of the board. In this case the edge of the board is terminated by a UPML boundary.

Excitation was provided by a current source implemented by introducing modified semi-explicit field update coefficients at the source location [6]. A differentiated Gaussian pulse excitation is used given by equation 8 with $J_0 = 1\ A/mm^2$, $t = 100\ pS$ and $\tau = 50\ pS$.

$$J_t = J_0(t - t_0)e^{-\left(\frac{t-t_0}{\tau}\right)} \tag{8}$$

Examination of the Fourier transform of the current pulse confirms it has significant components to frequencies > *15 GHz*.

The board dimensions are $L_x$ = 100 *mm,* $L_y$ = *60 mm* and $L_z$ = *0.8 mm* and the size of the Yee grid is $\Delta x = \Delta y = 2$ *mm* and $\Delta z = 0.4$ *mm.* Hence according to equation 6, dispersion should not be evident for frequencies up to *30 GHz*. A time step of *1 pS* was used which is well below the CFL stability limit of *1.33 pS* given by equation 7.

Adequate convergence using only 2 cells in the *z*-axis was confirmed by running a smaller model in *x* and *y* dimensions with 1 to 15 cells in the *z*-axis where no significant difference the electric field ( < 0.1 %) was seen.

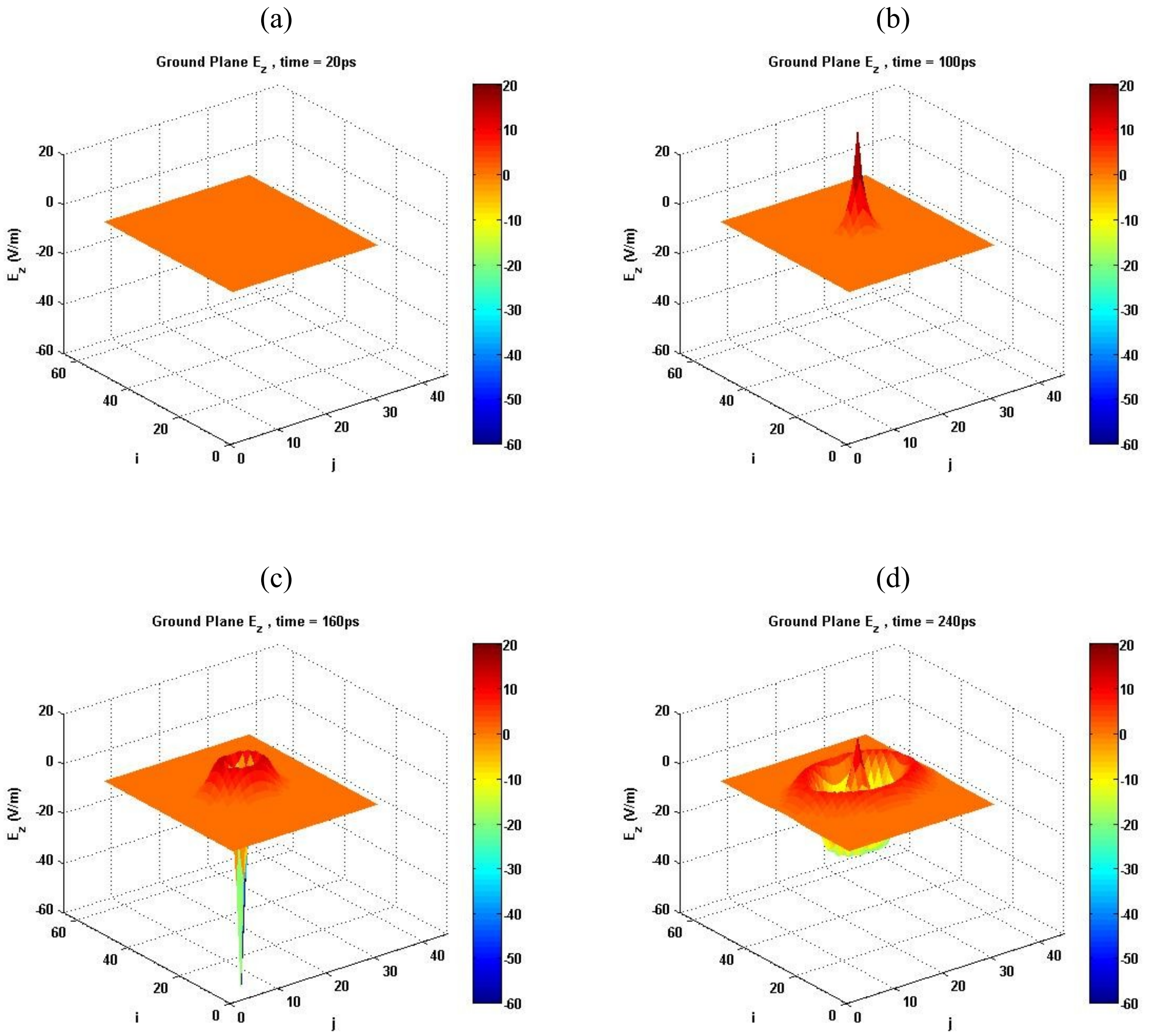

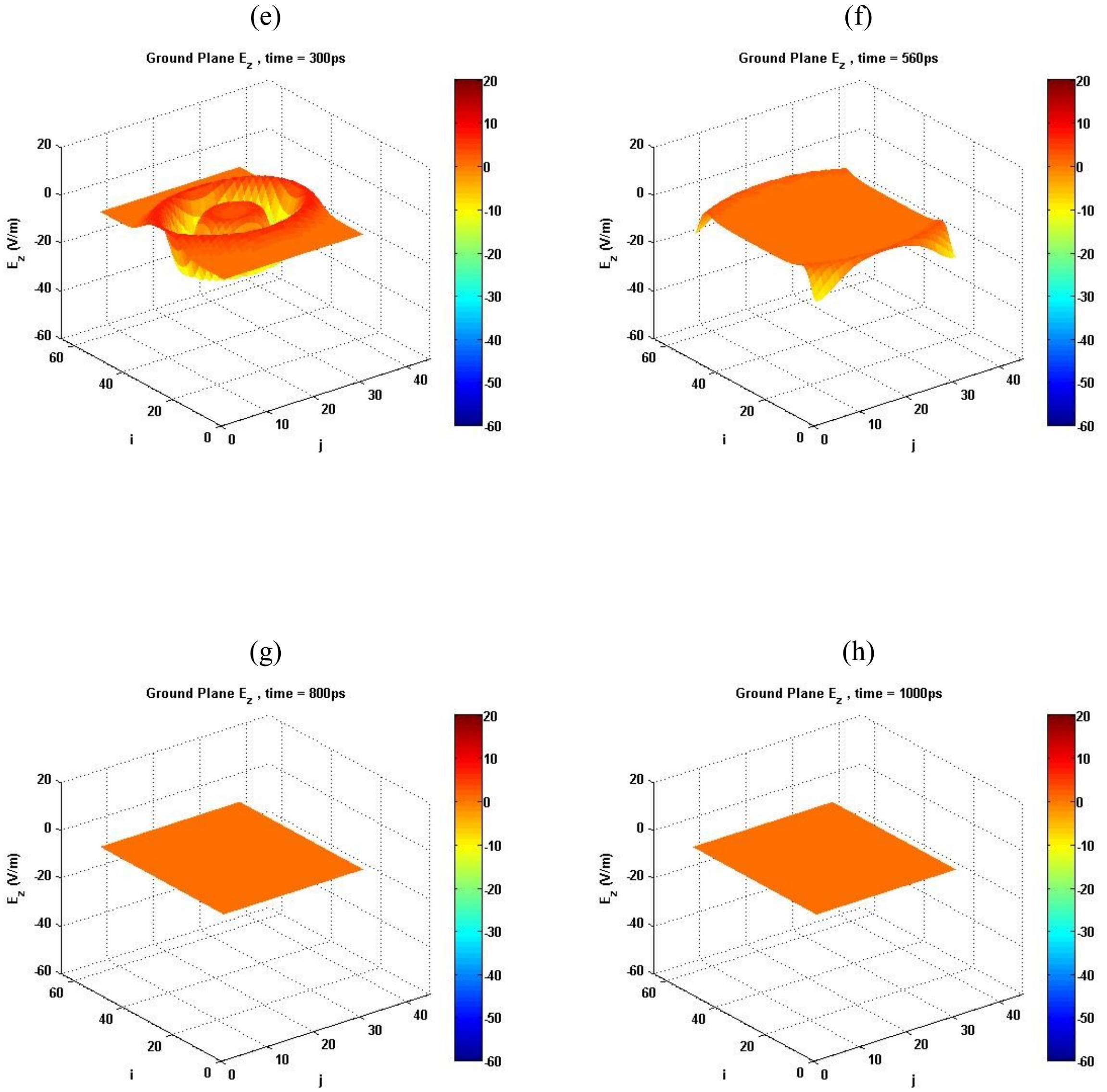

**Figure 3 Evolution of ground plane $E_z$ distribution with UPML boundary.**

From figure 3 it can be seen that the wave front spreads radially over the surface of the board and has decayed to zero by ~ *600 ps*. Figure 4 shows a similar model with an air layer included between the edge of the PCB and the UPML boundary. In this case the wave front is reflected from the dielectric/air interface in the *y*-axis after ~ *300 ps*. A complex interference pattern is set up between source and reflected wave fronts and then between multiply reflected wave fronts which affects the whole power and ground plane area and could adversely affect the operation of any device on the board. These resonant waves persist for a significant time after the initial pulse. Although it should be noted that this is overestimated as the model uses a lossless dielectric and perfect (i.e. lossless) electric conductors, although in practice these losses will be small at the frequencies of interest. The model does capture the small transmission loss which does occur as there is some field coupling across the dielectric/air boundary. This too is of interest as the radiated fields can result in electromagnetic interference [7]

(a)

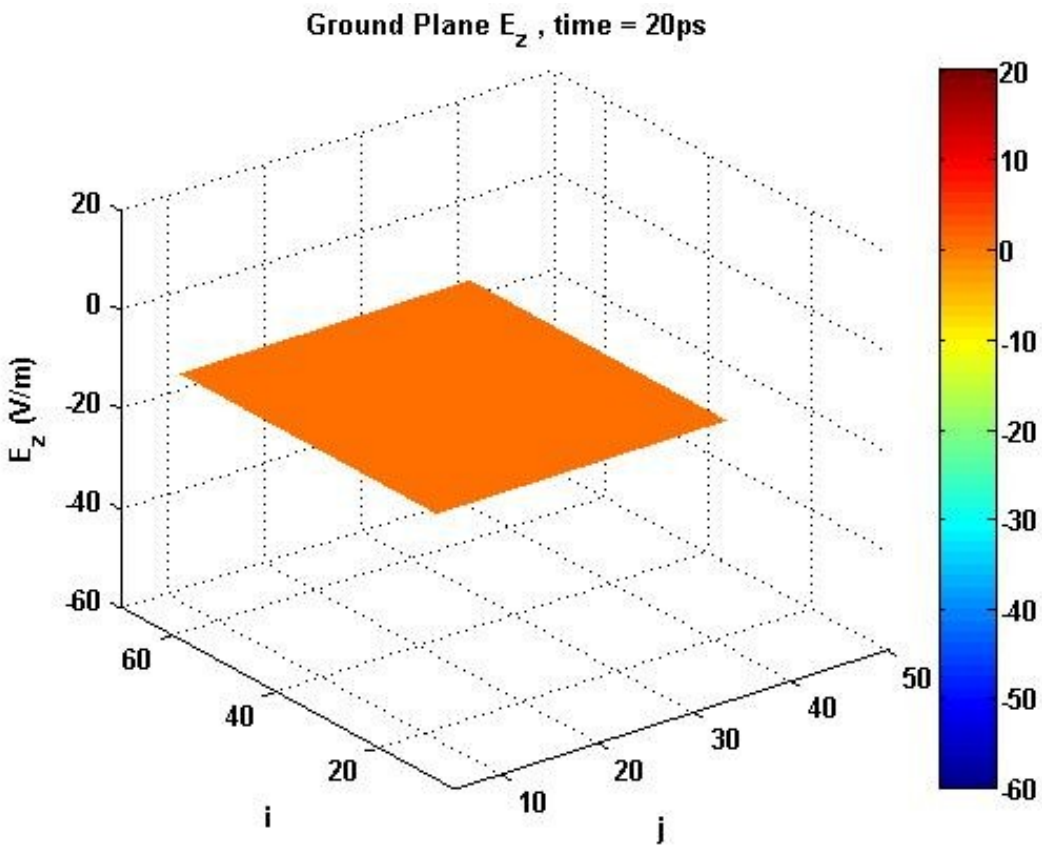

Ground Plane $E_z$ , time = 20ps
$E_z$ (V/m)
20
0
-20
-40
-60
60
40
20
i
10
20
30
40
50
j
20
10
0
-10
-20
-30
-40
-50
-60

(b)

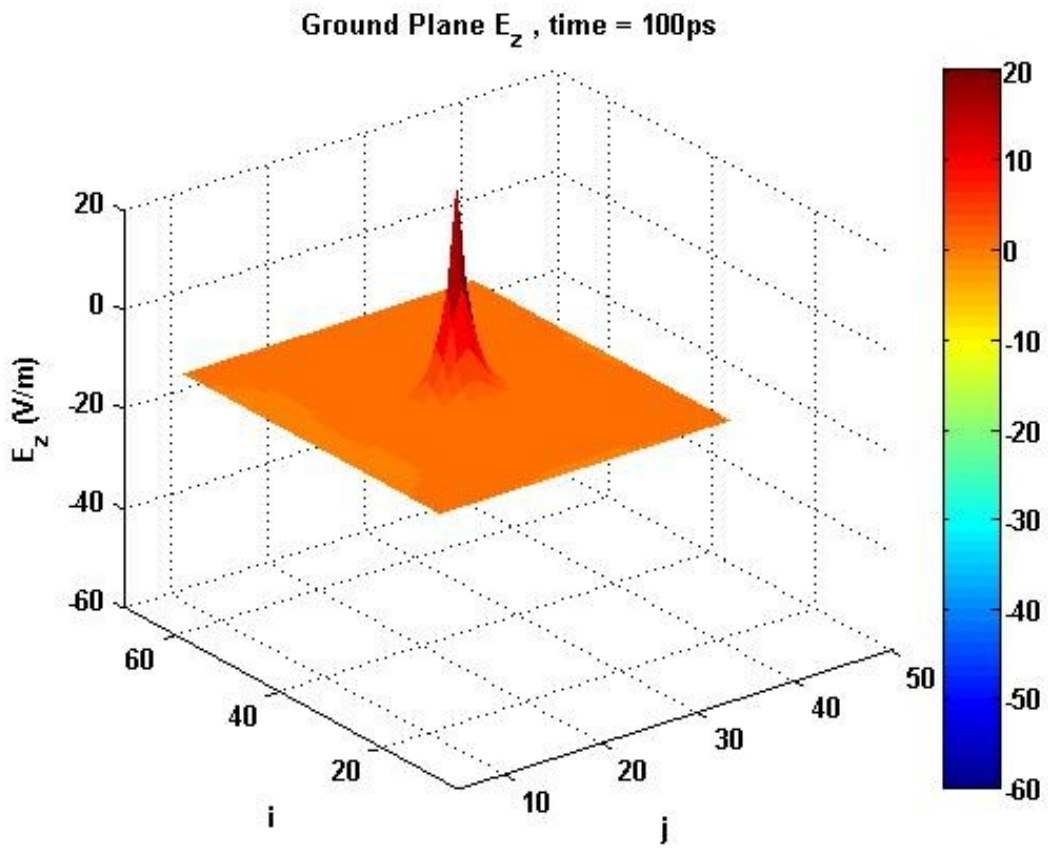

Ground Plane $E_z$ , time = 100ps
$E_z$ (V/m)
20
0
-20
-40
-60
60
40
20
i
10
20
30
40
50
j
20
10
0
-10
-20
-30
-40
-50
-60

(c)

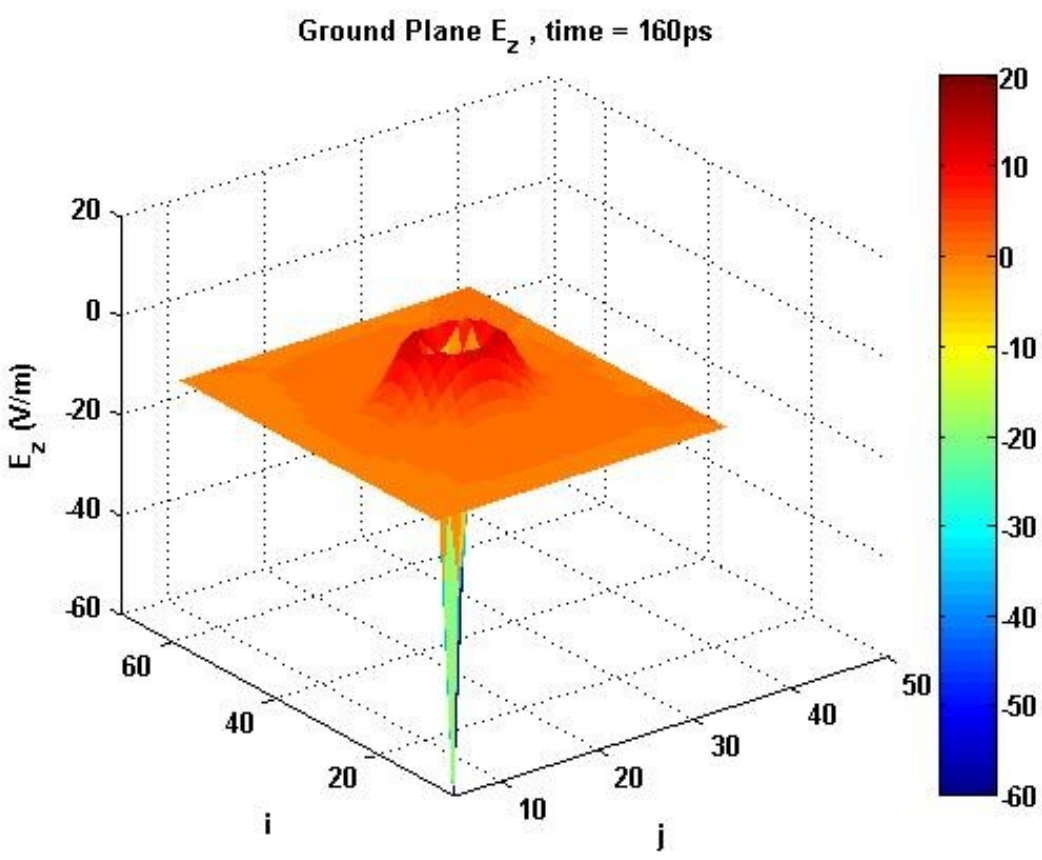

Ground Plane $E_z$ , time = 160ps
$E_z$ (V/m)
20
0
-20
-40
-60
60
40
20
i
10
20
30
40
50
j
20
10
0
-10
-20
-30
-40
-50
-60

(d)

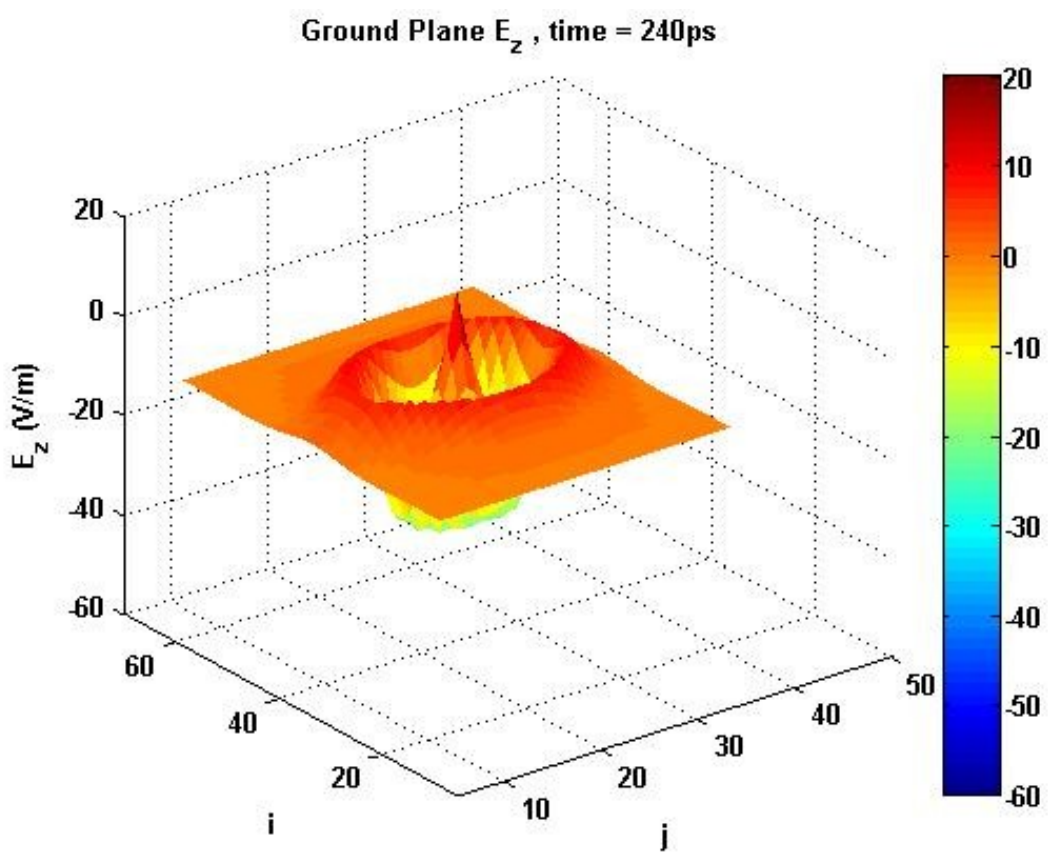

Ground Plane $E_z$ , time = 240ps
$E_z$ (V/m)
20
0
-20
-40
-60
60
40
20
i
10
20
30
40
50
j
20
10
0
-10
-20
-30
-40
-50
-60

(e)

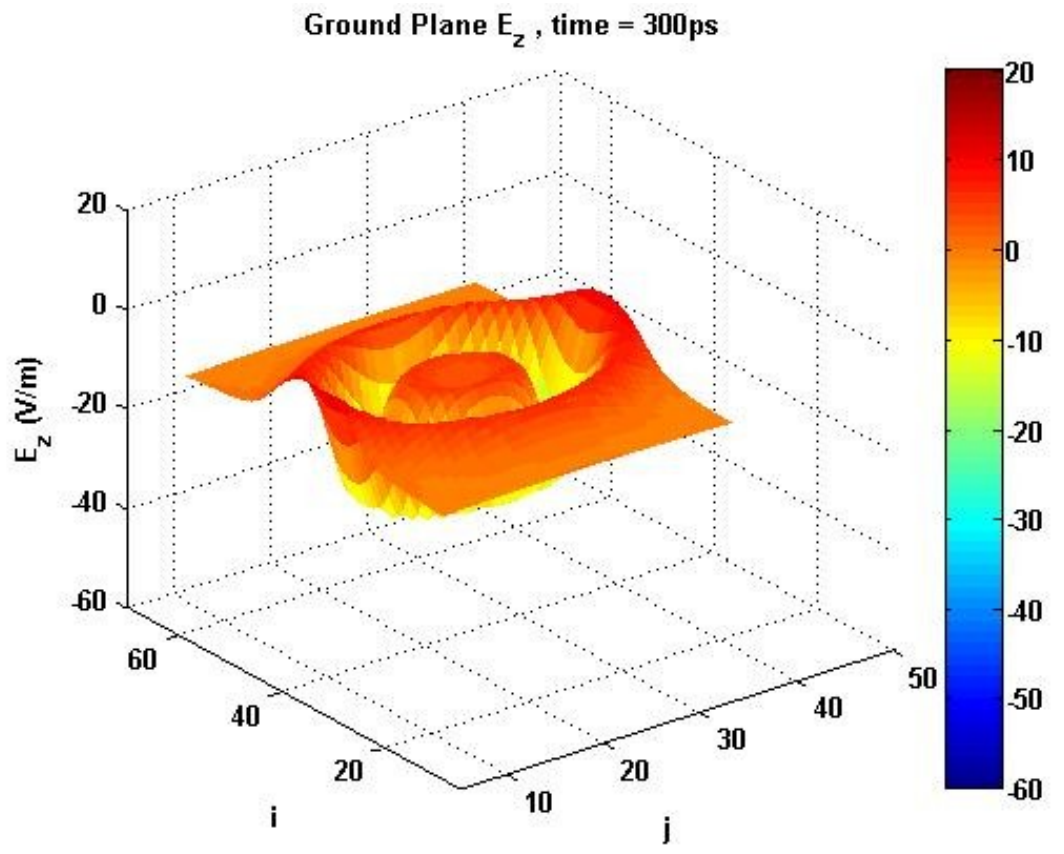

(f)

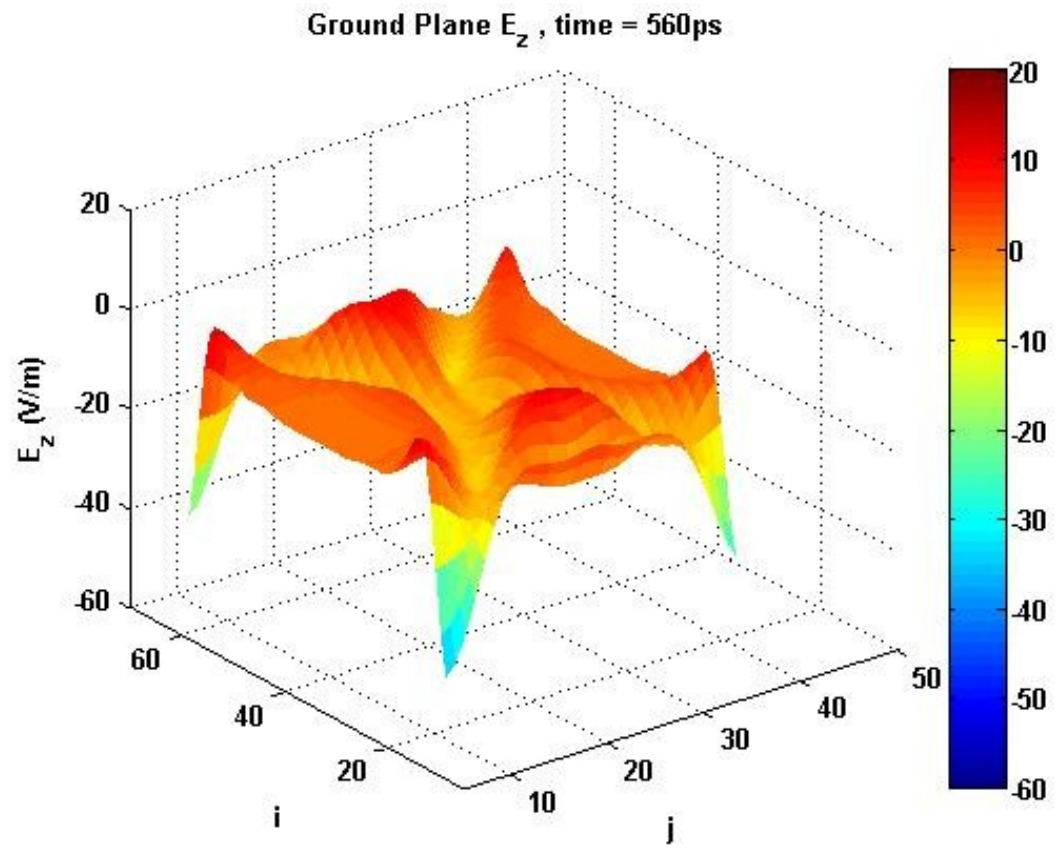

(g)

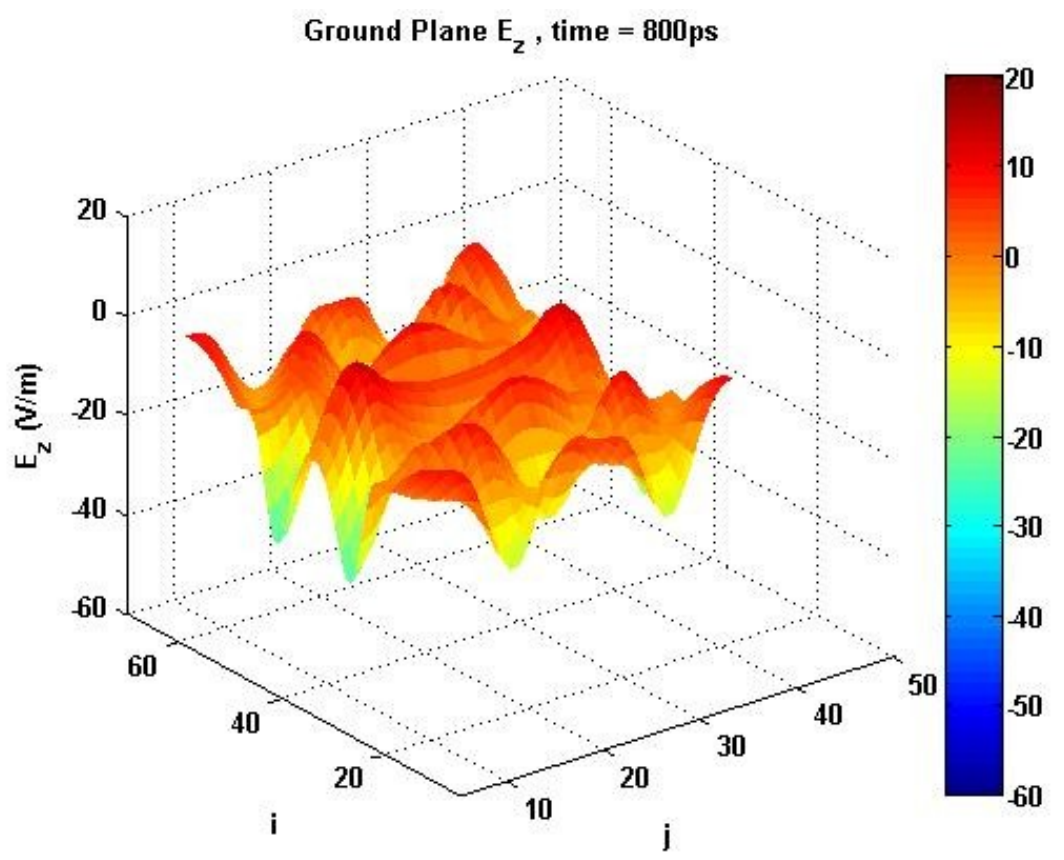

(h)

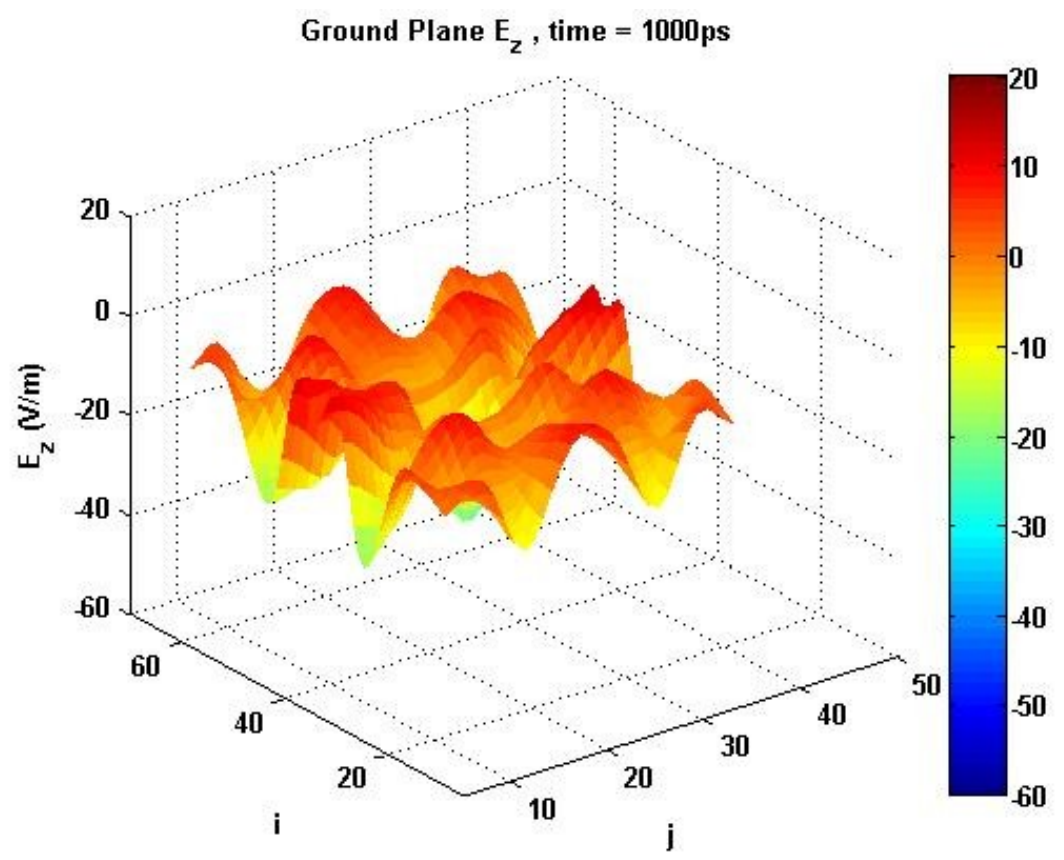

**Figure 4 Evolution of ground plane $E_z$ distribution for PCB surrounded by an air layer.**

The power planes can be modelled as a resonant cavity with two perfect electric conductor boundaries (corresponding to the power and ground planes) and four perfect magnetic conductor boundaries corresponding to the dielectric/air interfaces when the resonant frequencies are given by equation 9 [8, 9]

$$f_r = \frac{c}{\sqrt{\epsilon_r}}\sqrt{\left(\frac{m}{2\left(L_x + {L_z}/{2}\right)}\right)^2 + \left(\frac{n}{2\left(L_y + {L_z}/{2}\right)}\right)^2} \qquad (9)$$

where $c$ is the velocity of light, $\varepsilon_r$ is the relative permittivity, $m$ and $n$ are integers representing the order of the resonant mode and $L_x$,$L_y$ and $L_z$ are the board dimensions. Note that the small addition of $(L_z/2)$ to the $L_x$, and $L_y$ dimensions in equation 8 is intended to approximate the effects of fringing fields at the dielectric/air interfaces.

A fast Fourier transform (FFT) of the ground plane $E_z$ did not show the expected resonant behaviour presumably because the response is dominated by the source waveform. In addition it has been reported that not all modes are excited equally, there is some dependence on the location of the source and some modes can be suppressed depending on the location of the source and measurement ports [9]. To avoid any possible mode suppression the source was moved a corner of the board and the ground plane $E_z$ measured at the opposite corner. The current source as replaced by a resistive voltage source implemented using the appropriate semi-explicit update coefficients [6] with amplitude $V_0 = 5V$ and source resistance $R_S = 10\Omega$. This allows any effects due to source match or mismatch to be investigated by varying the source resistance. The voltage excitation used the same differentiated Gaussian pulse shape as for the current source. The board size was reduced to $L_x = 40$ *mm*, $L_y = 30$ *mm* and $L_z = 0.8$ *mm* to increase the resonant frequencies to allow them to be resolved in fewer time steps.

Figure 5 shows a FFT of the ground plane $E_z$ measured at the opposite corner to the source after the model was run for $3 \times 10^{-8}$s (30000 steps). The resonant peaks are generally well resolved. Table 1 compares the frequencies of the peaks with the values expected from equation 9. The measured and calculated frequencies are in excellent agreement. The modes labelled J and O are only just resolved and were assigned by closely examining the spectrum. However it should be noted that that the predicted frequency separation from their neighbours is small. Unfortunately it was not possible to improve the resolution by increasing the number of time steps further without exceeding the available memory of the 64 bit / 8GB PC.

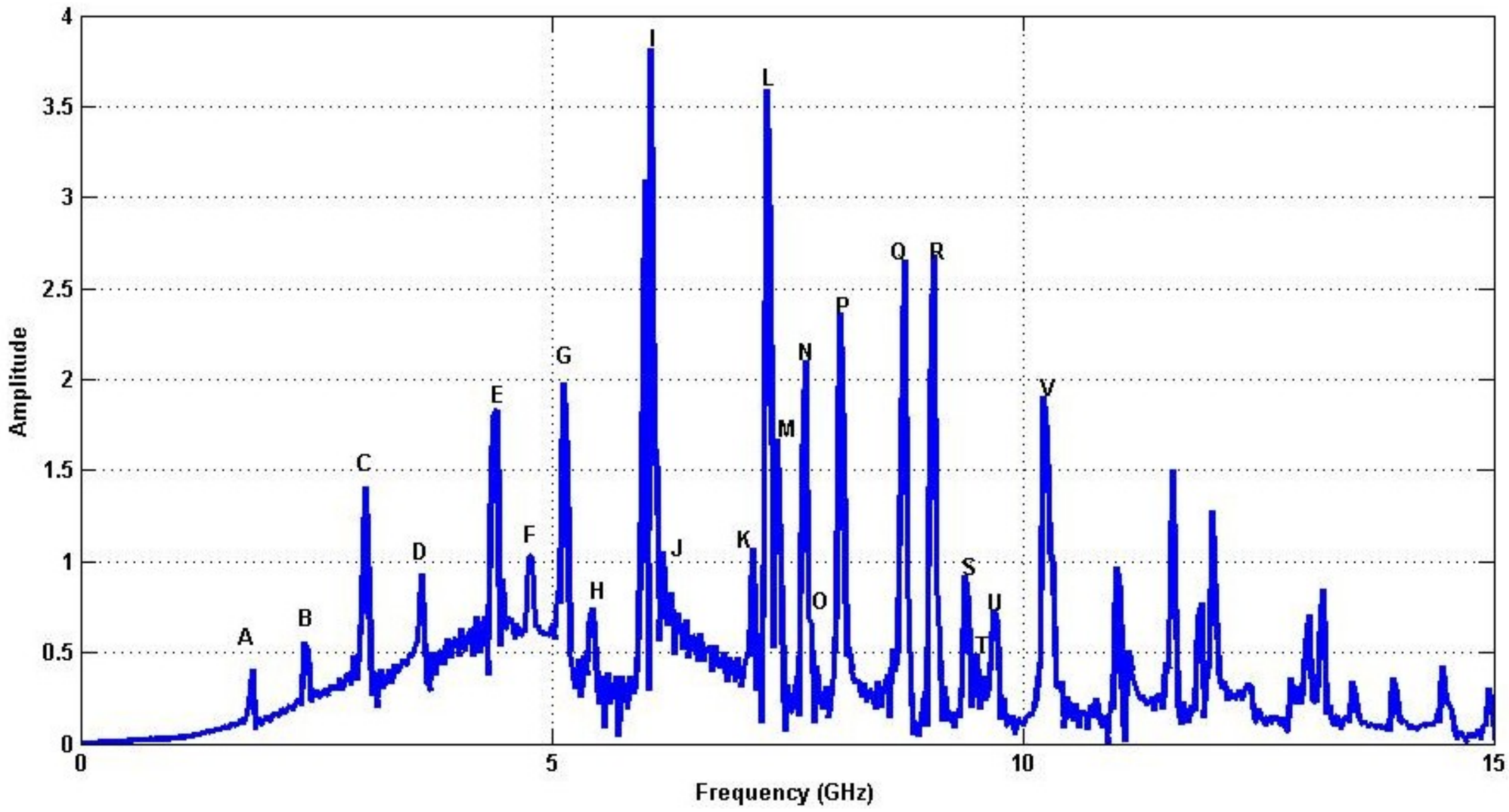

**Figure 5 Frequency spectrum of ground plane $E_z$ for board size $L_x$ = 40 mm, $L_y$ = 30 mm and $L_z$ = 0.8 mm.**

| *m* | *n* | *Predicted Frequency (GHz)* | *Label (Figure 5)* | *Modelled Frequency (GHz)* |
|---|---|---|---|---|
| **1** | **0** | **1.86** | **A** | **1.82** |
| **0** | **1** | **2.47** | **B** | **2.38** |
| **1** | **1** | **3.09** | **C** | **3.03** |
| **2** | **0** | **3.71** | **D** | **3.62** |
| **2** | **1** | **4.46** | **E** | **4.42** |
| **0** | **2** | **4.93** | **F** | **4.79** |
| **1** | **2** | **5.27** | **G** | **5.13** |
| **3** | **0** | **5.57** | **H** | **5.44** |
| **3** | **1** | **6.09** | **I** | **6.06** |
| **2** | **2** | **6.18** | **J** | **6.18** |
| **0** | **3** | **7.40** | **K** | **7.14** |
| **4** | **0** | **7.43** | **L** | **7.30** |
| **3** | **2** | **7.44** | **M** | **7.38** |
| **1** | **3** | **7.63** | **N** | **7.69** |
| **4** | **1** | **7.82** | **O** | **7.75** |
| **2** | **3** | **8.28** | **P** | **8.16** |
| **4** | **2** | **8.92** | **Q** | **8.74** |
| **3** | **3** | **9.26** | **R** | **9.05** |
| **5** | **0** | **9.28** | **S** | **9.39** |
| **5** | **1** | **9.60** | **T** | **9.52** |
| **0** | **4** | **9.87** | **U** | **9.70** |
| **1** | **4** | **10.04** | **V** | **10.23** |

**Table 1 Comparison of measured and calculated resonant frequencies for board size $L_x$ = 40 mm, $L_y$ = 30 mm and $L_z$ = 0.8 mm.**

## Conclusion

Power integrity is becoming increasingly relevant due to increases in device functionality and switching speeds along with reduced operating voltage. Large current spikes at the device terminals result in electromagnetic disturbances which can establish resonant patterns affecting the operation of the whole system.

These effects have been examined using a finite difference time domain approach to solve Maxwell's equations for the PCB power and ground plane configuration. The simulation domain is terminated with a uniaxial perfectly matched layer to prevent unwanted reflections. This approach calculates the field values as a function of position and time and allows the evolution of the field to be visualized.

The propagation of a pulse over the ground plane was observed demonstrating the establishment of a complex interference pattern between source and reflected wave fronts and then between multiply reflected wave fronts affecting the whole ground plane area. This wave interference effect could adversely affect the operation of any device on the board. These resonant waves persist for a significant time after the initial pulse. Examining the FFT of the ground plane electric field response showed numerous resonant peaks at frequencies in excellent agreement with the expected values assuming the PCB can be modelled as a resonant cavity with two electric and four magnetic field boundaries.